\documentclass[sigplan,10pt]{acmart}
\acmConference[PPoPP'26]{ACM SIGPLAN Annual Symposium on Principles and Practice of Parallel Programming}{January 31--February 04, 2026}{Sydney, Australia}
\acmYear{2026}
\acmISBN{} 
\acmDOI{} 
\startPage{1}

\setcopyright{none}

\usepackage{booktabs}   
\usepackage{subcaption} 

\usepackage{makecell}

\begin{document}

\title[PPoPP Paper]{High-Performance N-Queens Solver on GPU: Iterative DFS with Zero Bank Conflicts}         


 \author{Guangchao Yao}
 \affiliation{
   \institution{XiaoPeng Motors}            
   \city{Beijing}
   \country{China}                    
 }
 \email{yaogc@xiaopeng.com}          

 \author{Yali Li}
 \affiliation{
   \institution{XiaoPeng Motors}           
   \city{Beijing}
   \country{China}                   
}
\email{liyl19@xiaopeng.com}         

\fancyhead{}  
\renewcommand\footnotetextcopyrightpermission[1]{} 

\begin{abstract}
The counting of solutions to the N-Queens problem is a classic NP-complete problem with extremely high computational complexity. As of now, the academic community has rigorously verified the number of solutions only up to N $\leq$ 26. In 2016, the research team led by Preußer solved the 27-Queens problem using FPGA hardware, which took approximately one year, though the result remains unverified independently. Recent studies on GPU parallel computing suggest that verifying the 27-Queens solution would still require about 17 months, indicating excessively high time and computational resource costs. To address this challenge, we propose an innovative parallel computing method on NVIDIA GPU platform, with the following core contributions: (1) An iterative depth-first search (DFS) algorithm for solving the N-Queens problem; (2) Complete mapping of the required stack structure to GPU shared memory; (3) Effective avoidance of bank conflicts through meticulously designed memory access patterns; (4) Various optimization techniques are employed to achieve optimal performance. Under the proposed optimization framework, we successfully verified the 27-Queens problem in just 28.4 days using eight RTX 5090 GPUs, thereby confirming the correctness of Preußer's computational results. Moreover, we have reduced the projected solving time for the next open case—the 28-Queens problem—to approximately 11 months, making its resolution computationally feasible. Compared to the state-of-the-art GPU methods, our method achieves over 10× speedup on identical hardware configurations (8 A100), while delivering over 26× acceleration when utilizing 8 RTX 5090 GPUs, and brings fresh perspectives to this long-stagnant problem\footnote{Code can be found at https://github.com/ygch/n\_queens.}. 
\end{abstract}

\maketitle

\section{Introduction}
The N-Queens problem is a classical puzzle originating from the historic 8-Queens problem \cite{ref1}. The objective is to place N queens on an N×N chessboard such that no two queens threaten each other horizontally, vertically, or diagonally. While finding a single valid solution is relatively straightforward, precisely counting all valid solutions dramatically increases computational complexity. Currently, no general closed-form solution exists, leaving exhaustive search with backtracking as the only viable approach. 

The most efficient classical backtracking algorithm is the Somers algorithm \cite{ref2}, which employs bitwise operations to rapidly eliminate invalid positions, thereby dramatically reducing the search space. Furthermore, its non-recursive implementation offers excellent portability, making it the foundation for many parallel computing implementations. However, even the Somers algorithm reaches its computational limits when dealing with large-scale N-Queens problems.

The N-Queens problem can be decomposed into a set of independent subproblems by pre-placing queens in the first few rows. These subproblems can then be solved in parallel, with their results aggregated for the final count (Figure 1). During pre-placement, only the first half of the positions in the first row need to be considered, as the solutions for the latter half are symmetric. Additionally, if N is odd, placing a queen in the central position of the first row allows symmetry-based optimizations for the second row. Due to symmetry—only N/2 - 1 positions need evaluation\footnote{The subtraction accounts for the three central positions in the second row being blocked by the queen in the first row.}. As shown in the figure, all valid solutions to the N-Queens problem form a tree, where solving the problem equates to traversing this tree. Each subproblem corresponds to an independent subtree, with no overlap between subtrees. This inherent independence provides ideal conditions for parallel acceleration, enabling subproblems to be solved entirely in parallel.

\begin{figure*}[t]
	\centering
	\includegraphics[width=0.7\textwidth]{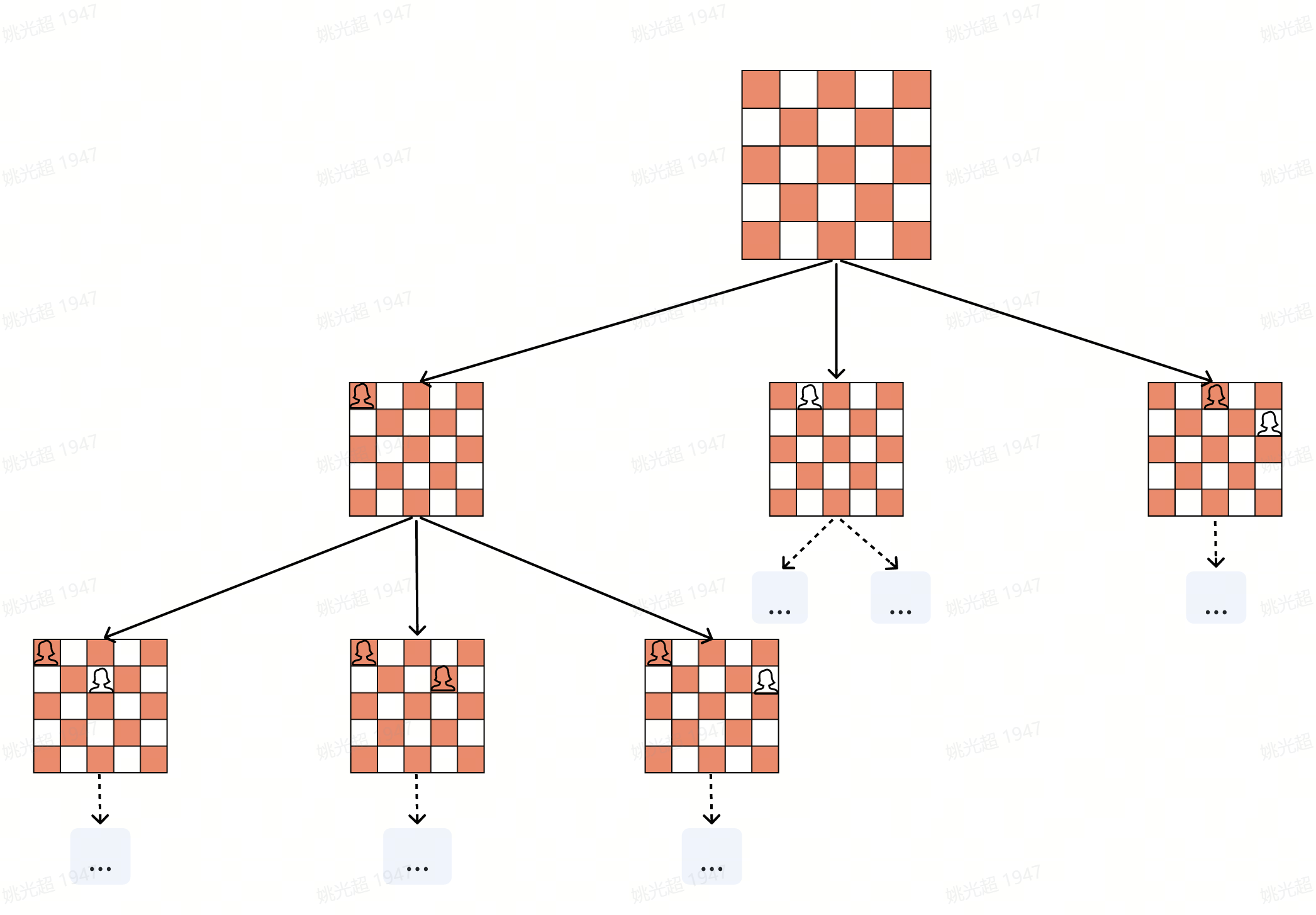}
	\caption{First two-row expansion in the 5-Queens problem. Aggregate the number of leaf nodes in each subtree to obtain the final result.} 
\end{figure*}

\subsection{Related Work}
Three primary parallel optimization approaches are commonly employed: distributed multi-core computing, FPGA, and GPU. Distributed multi-core computing approach typically involves partitioning subproblems across multiple nodes and utilizing numerous CPU cores for computation. In 2004, the Kise’s team \cite{ref3} successfully solved the 24-Queens problem using 68 Pentium4 Xeon processors in 22 days. Subsequently in 2007, Caromel's team \cite{ref4} employed 260 machines to solve the 25-Queens problem over 185 days. Both implementations relied on extensive CPU core clusters for parallel processing. However, as general-purpose processors, CPUs demonstrate limited advantages in parallel computation.

FPGAs later emerged as a breakthrough technology due to their high parallelism and hardware programmability. In 2009, Preußer et al. \cite{ref5} solved the 26-Queens problem in 9 months using FPGA. In 2017, the same team calculated the solution count for the 27-Queens problem \cite{ref6} in about one year. However, due to the substantial computational resources and time required for cross-verification, the solution for the 27-Queens problem currently remains unverified \cite{ref7}.

With the rise of GPU as a general-purpose parallel computing platform, the N-Queens problem has become an extensively studied benchmark, leading to numerous GPU-based parallel implementations. Early work by \cite{ref8} used an NVIDIA GTX 480 GPU to reduce the solving time for the 19-Queens problem to 50 seconds. Thouti et al. \cite{ref9} implemented GPU parallelization via OpenCL, achieving a 20× speedup for N = 16–21. The latest advancement by \cite{ref10} ported the DoubleSweep-Light algorithm to GPUs, reducing the 19-Queens solving time to 1.5 seconds and the 24-Queens solving time to 16 hours using 8 NVIDIA A100 GPUs. Based on their performance metrics, they estimated that verifying the 27-Queens solution could be accomplished within 17 months \footnote{This estimate is overly optimistic; the actual time required is likely to exceed two years.}. Despite these remarkable speed improvements, current computational capabilities remain insufficient for tackling the unsolved 28-Queens problem. The exponential growth in solution space continues to present formidable challenges, even with modern parallel computing architectures.

\subsection{Fundamentals of GPU Computing}
Since NVIDIA introduced the CUDA programming language, GPUs have gradually emerged as a dominant platform for parallel computing. The advent of CUDA has enabled developers to more efficiently harness the substantial computational power of GPUs, significantly advancing research progress on the N-Queens problem. The primary advantage of GPU lies in its massive number of computational cores and exceptional memory bandwidth, which dramatically accelerates the processing of complex problems. For instance, the RTX 4090 GPU features 16,384 CUDA cores. If these resources can be fully utilized, the solving speed for the N-Queens problem could theoretically reach unprecedented levels.

A critical factor in achieving performance acceleration on GPUs lies in the effective utilization of shared memory. As an on-chip memory resource, shared memory exhibits significantly lower access latency (typically tens of clock cycles) compared to global memory, which resides off-chip and suffers from substantially higher latency (ranging from hundreds to thousands of clock cycles). Proper exploitation of shared memory can therefore dramatically reduce data access latency and improve overall computational efficiency. The availability of shared memory represents a distinctive advantage of GPU programming, as it provides developers with direct control over on-chip cache - a capability generally unavailable in CPU architectures where on-chip cache access is not directly programmable. This architectural feature enables fine-grained optimization of memory access patterns that is crucial for memory-intensive problems like the N-Queens problem.

Although shared memory features extremely low access latency, improper usage may still result in suboptimal performance gains. From a hardware perspective, shared memory is partitioned into 32 banks, each with a 32-bit width, where addresses increment sequentially across banks as illustrated in Figure 2. Special attention must be paid to avoiding bank conflicts. When threads within a warp access shared memory, high bandwidth can be achieved if each thread accesses data from different banks, enabling parallel processing of all requests. However, if multiple threads access different addresses within the same bank, bank conflicts occur, forcing serialized access. In such cases, the hardware must split these accesses into multiple sequentially executed requests, consequently increasing latency. To fully exploit the low-latency advantage of shared memory, it is crucial to carefully design data structures and access patterns to ensure threads access different banks.

\begin{figure}[H]
	\centering
	\includegraphics[width=8cm]{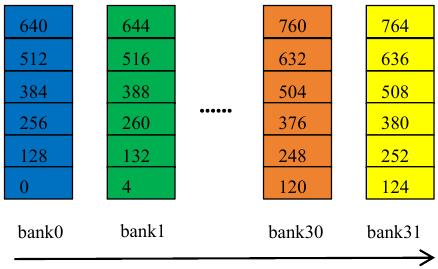}
	\caption{Shared memory bank partitioning and address mapping} 
\end{figure}

Another critical consideration when using shared memory is its limited capacity, typically only tens of kilobytes \cite{ref11}. Thus, algorithms must be carefully designed to prevent data overflow. For the N-Queens problem, as N increases, the required stack space also grows, often exceeding the limit of shared memory size. This necessitates a trade-off between time and space efficiency to optimize performance.

\section{Iterative Depth-First Search N-Queens}
This section begins with an overview of three approaches for the N-Queens problem based on Somers algorithm. Somers algorithm employs state compression techniques and bitwise operation optimization, significantly reducing memory space requirements. The algorithm utilizes three 32-bit variables (\textbf{cur, left, and right}) to encode the occupied states of columns, diagonals, and anti-diagonals respectively, as illustrated in figure 3. 

\begin{figure}[H]
	\centering
	\includegraphics[width=5cm]{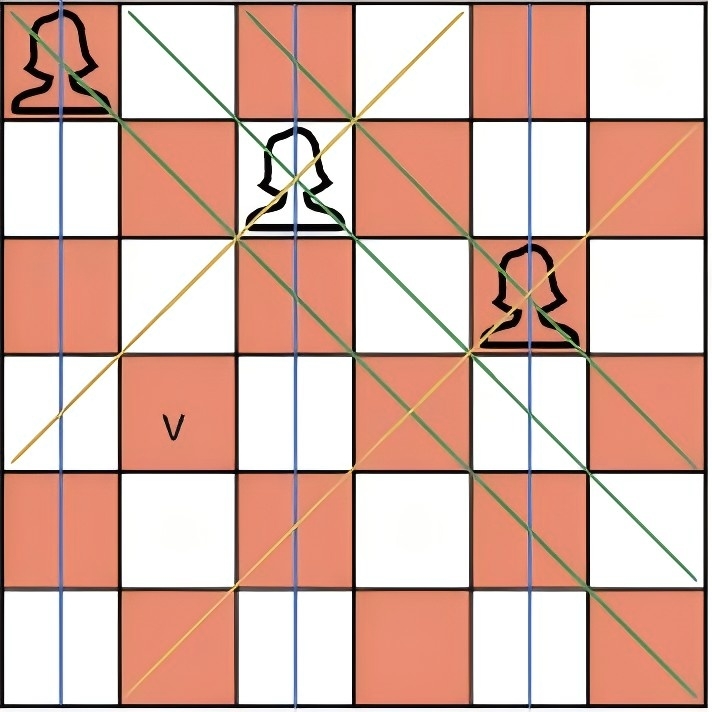}
	\caption{Somers algorithm uses cur(blue), left(yellow), and right(green) to efficiently filter out invalid positions in 6-queens problem} 
\end{figure}

When a queen is placed at a valid position p in a new row, the corresponding column becomes prohibited. This column constraint is enforced by performing a bitwise OR operation between p and cur, effectively marking the column as occupied. Similarly, to restrict the left-diagonal attacks in the next row, a bitwise OR is applied between p and left, followed by a left shift by one bit. This propagates the left-diagonal threat along the $\swarrow$ direction. Likewise, for the right-diagonal constraints, a bitwise OR between p and right is performed, followed by a right shift by one bit. This propagates the right-diagonal threat along the $\searrow$ direction. As for finding the valid positions in the next row, we only need to compute the bitwise OR of cur, left, and right, and then take the bitwise NOT to obtain all valid positions. The original Somers algorithm adopts a non-recursive implementation, which presents considerable comprehension challenges. To facilitate understanding, we first present a simplified recursive implementation shown in algorithm 1.

\begin{table}[H]
	\centering
	\begin{tabular}{l}
		\toprule
		\textbf{Algorithm 1} Recursive Somers Algorithm\\ 
		\midrule
1: \textbf{def}  n\_queens(N, cur, left, right, sum):  \\
2: \quad   last = (1 $\ll$ N) - 1  \\
3: \quad     \textbf{if} cur == last:  \\
4: \qquad        sum += 1  \\
5: \qquad        return  \\
6: \quad     valid\_pos = last \& ( $\sim$(cur | left | right))  \\
7: \quad     \textbf{while} valid\_pos != 0:  \\
8: \qquad        p = valid\_pos \& (-valid\_pos)  \\
9: \qquad        valid\_pos -= p  \\
10:\quad \space\space\space    cur = cur | p\\
11:\quad \space\space\space    left = (left | p) $\ll$ 1\\
12:\quad \space\space\space    right = (right | p) $\gg$ 1\\
13:\quad \space\space\space    n\_queens(N, cur, left, right, sum)\\
		\bottomrule
	\end{tabular}
\end{table}

We employ algorithm 1 to generate a series of subproblems by pre-placing queens in the first k rows. Specifically, starting from the first row, all feasible positions are traversed(only the first half). After k iterations, the placement of k queens is completed, subsequently generating a series of (cur, left, right) triplets. Each triplet represents a distinct subproblem corresponding to a partial solution state. Although all subproblems share the same recursion depth, the size of their solution spaces exhibits considerable variation. This, in turn, leads to significant differences in the time required to solve each individual subproblem. This was confirmed by subsequent Multi-GPU evaluation parts.

However, the recursively implemented algorithm 1 is not CUDA-friendly and must be transformed into an iterative form by explicitly simulating the recursion call using a stack structure. As previously mentioned, the solution space for each subproblem forms a subtree, where each node represents the placement of queens in the first \textit{k} rows, and each branch denotes a valid placement position for the \textit{k+1} row. For traversing such trees, there are generally two approaches: breadth-first search (BFS) and depth-first search (DFS). BFS resembles level-order traversal of a node, where all branches of a node are pushed onto the stack at once. Algorithm 1 is just a BFS-based implementation. In contrast, DFS explores only one branch at a time, proceeding along a single path until reaching its deepest point, and then backtracking to the previous node to continue exploration.

From the perspective of space complexity, DFS demonstrates significant advantages. The maximum depth of its stack space is strictly equal to N minus the number of pre-placed rows, and this deterministic upper bound ensures predictable memory requirements. Whenever the search reaches the maximum depth, the algorithm either successfully finds a valid solution or performs backtracking to return to the previous level and continue the search.

In contrast, BFS must simultaneously maintain all possible branches of the current node, resulting in a worst-case space complexity of O(N×d). Theoretically, this imposes an N-fold increase in memory demand compared to DFS (though in practice, this gap is often somewhat reduced). Given the scarcity of GPU shared memory resources (typically limited to just tens of KB), adopting a DFS strategy minimizes memory usage, potentially allowing the complete stack structure to reside entirely within shared memory. This characteristic makes DFS the ideal choice for GPU implementations.

Therefore, our CUDA-based N-Queens algorithm employs a DFS strategy. It should be noted that the DFS implementation requires additional storage of the \textbf{valid\_pos} variable in the stack space. Nevertheless, its overall memory consumption remains significantly lower than that of the BFS approach. The code for the iterative N-Queens algorithm based on depth-first search is shown in Algorithm 2.

\begin{table}[H]
	\centering
	\begin{tabular}{l}
		\toprule
		\textbf{Algorithm 2}  DFS-based Iterative Somers Algorithm\\ 
		\midrule
1: \textbf{def}  n\_queens\_iterative(N, cur, left, right, sum):\\
2:\quad      last = (1 $\ll$ N) - 1\\
3:\quad      stack = [0] * 96\\
4:\quad      top = 0\\
5:\quad      valid\_pos = last \& ($\sim$(cur | left | right))\\
6:\quad      stack[top : top + 4] = [cur, left, right, valid\_pos]\\
7:\quad      top += 4\\
8:\quad      \textbf{while} top != 0:\\
9: \qquad         [cur, left, right, valid\_pos] = stack[top - 4 : top]\\
10:\qquad         p = valid\_pos \& (-valid\_pos)\\
11:\qquad         valid\_pos -= p\\
12:\qquad         stack[top - 1] = valid\_pos \\
13:\qquad         top -= (valid\_pos == 0 ? 4 : 0)\\
14:\qquad         cur = cur | p\\
15:\qquad         \textbf{if} cur == last: \\
16:\qquad   \quad          sum += 1\\
17:\qquad   \quad          continue\\
18:\qquad         left = (left | p) $\ll$ 1\\
19:\qquad         right = (right | p) $\gg$ 1\\
20:\qquad         valid\_pos = last \& ($\sim$(cur | left | right))\\
21:\qquad         \textbf{if} valid\_pos == 0:\\
22:\qquad  \quad           continue\\
23:\qquad         stack[top : top + 4] = [cur, left, right, valid\_pos]\\
24:\qquad         top += 4\\
		\bottomrule
	\end{tabular}
\end{table}

We have designed a stack structure with a capacity of 96 bytes \footnote{ A value corresponding to the maximum shared memory size per thread under static allocation strategy in NVIDIA GPUs such as the RTX 4090.} in algorithm 2. This stack is used to store four critical state variables: cur, left, right, and valid\_pos, with a maximum supported depth of 24 layers. When N $\leq$ 24, it can handle cases where no initial rows are pre-placed. If N > 24, we can pre-place the first few rows to process subproblems with reduced depth. To fully leverage the GPU's massive parallel computing capability, it is essential to generate a sufficient number of independent subproblems through row pre-placement strategies. Experimental results (detailed in Section 4.1) demonstrate that pre-placing 5–6 rows yields optimal parallel performance in GPU implementations. When pre-placing 5 rows, algorithm 2 can be used to solve the 28-Queens problem.

Compared to the original Somers algorithm, the above iterative depth-first search implementation of the N-Queens problem is more concise. By processing subproblems, it reduces the required stack space size, making it feasible to entirely place the stack in the GPU's shared memory.

\section{Bank Conflict-Free GPU Implementation}
\subsection{Bank Conflict-Free Shared Memory Access}
When porting the iterative DFS-based N-Queens algorithm to GPU platforms, the code in algorithm 2 can be directly reused without modification. However, it should be noted that by default the stack variable resides in local memory, which is essentially part of global memory and suffers from high access latency. We can use sentence “\textbf{\_\_shared\_\_ int stack[48*256]}” to  transfer the stack from local memory to shared memory with size 48KB. This allocation size corresponds to the maximum static shared memory capacity natively supported by NVIDIA GPU architectures. For scenarios requiring shared memory exceeding 48KB, a dynamic allocation strategy must be employed, mentioned in Section 3.2.

Although migrating the stack from local memory to shared memory is relatively straightforward, the primary challenge lies in effectively avoiding shared memory bank conflicts. The key principle is to ensure each thread within a warp accesses only a dedicated bank. Based on this concept, we designed a shared memory access pattern as illustrated in Figure 4. Rather than being allocated in contiguous memory space, each thread's stack is spaced at 128-byte intervals. This arrangement guarantees that memory addresses accessed by all threads within a warp reside in distinct banks, effectively eliminating bank conflicts. Through this design, shared memory access speed is significantly enhanced, thereby dramatically improving the GPU's computational efficiency in solving the N-Queens problem.

\begin{figure}[H]
	\centering
	\includegraphics[width=8cm]{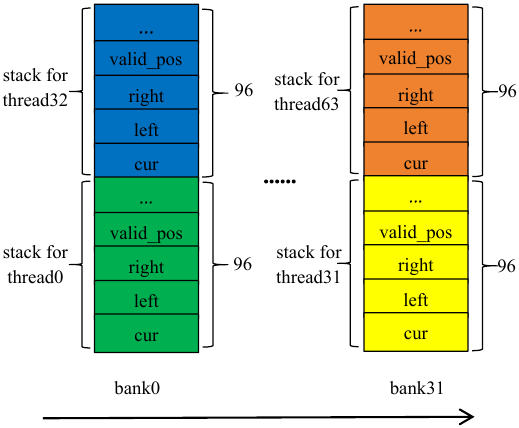}
	\caption{Stack space allocation in shared memory for each thread without bank conflict, each stack can hold 96 integers and support a maximum iteration depth of 24.} 
\end{figure}

The above shared memory access pattern can indeed avoid bank conflict, and deliver significant acceleration. However, to access cur, left, right, and valid\_pos, the GPU still needs to trigger four separate memory transactions. In practice, we can alternatively utilize CUDA's \textbf{int4} data type. Theoretically, using int4 would place these four variables in four consecutive memory banks, potentially causing 4-way bank conflicts within a warp. However, when using int4 types, CUDA automatically reduces the scheduling granularity from warp to quarter-warp, thereby eliminating bank conflicts. Using int4 type can also complete the access through four memory transactions.

Empirical results demonstrate that compared to the bank-conflict-free implementation mentioned above, the int4 approach provides additional performance gains, particularly when implemented with inline PTX assembly. Consequently, we adopt int4 as our default implementation.

\subsection{Optimization Techniques for Limited Shared Memory Capacity}
Due to the constraints of static shared memory allocation (maximum 48 KB), when each thread requires a stack space of 96 x 4 bytes, the number of threads per block must be limited to 128 (since 48 KB = 128 x 96 x 4 bytes). This configuration reduces per-block task throughput, thereby negatively impacting overall parallel computing efficiency. To address this issue, we propose three feasible optimization approaches.

\subsubsection{Approach 1}
To optimize parallel processing performance, we increase the number of threads per block within the 48KB shared memory constraint by reducing the stack space allocation per thread (while simultaneously decreasing the maximum recursion depth). For N-Queens problems of different scales, we designed multiple experimental configurations (Table 1). Given a specific N, we select Configurations 1, 2, or 3 according to their respective maximum supported values of N. Configurations 4 and 5 are excluded from subsequent experiments due to their limited applicability to small problem sizes, but are included here for completeness.

\begin{table}[H]
	\centering
	\caption{Different Thread Block and Stack Size Configurations}
\resizebox{\columnwidth}{!}{ 
	\begin{tabular}{|c |c| c|c|c|}
\hline
		\textbf{Config} & \textbf{\makecell{Block \\ Size}} & \textbf{\makecell{Stack\\ Size}} &   \textbf{\makecell{Recursion \\ Depth}} &  \textbf{\makecell{Max N \\ (Pre-placing first \\6 rows)}}\\ 
\hline
		1 & 128 & 96 & 24 & 30 \\ 
\hline
		2 & 160 & 76 & 19 & 25 \\ 
\hline
		3 & 192 & 64 & 16 & 22 \\
\hline
		4 & 256 & 48 & 12 & 18 \\
\hline
		5 & 512 & 24 & 6  & 12 \\
\hline
	\end{tabular}
}
\end{table}

\subsubsection{Approach 2}
This approach replaces static shared memory allocation with dynamic allocation to overcome the 48KB limitation. For GPUs like the A100 and RTX 4090, dynamic allocation allows expanding the shared memory per block to 96KB. However, at the hardware level, shared memory and L1 cache share the same physical resources. Increasing shared memory allocation consequently reduces L1 cache space, which may offset potential speed gains. This trade-off is experimentally validated in later sections. For comparative purpose, we designate this optimization as Configuration 6.

\subsubsection{Approach 3}
In Algorithm 1\&2, the recursion depth always reaches N levels if the solution is valid. However, when traversing to the (N-1)th row, we have already computed the valid positions (valid\_pos) for the Nth row. If valid\_pos equals 0, it indicates no feasible solution exists; otherwise, one valid solution is confirmed. This observation demonstrates that after completing the traversal of the (N-1)th row, no additional stack push operation is required - only a check of valid\_pos is needed to determine the result. Based on this optimization, we successfully reduce the maximum recursion depth by one level, consequently increasing the maximum solvable problem size N by 1 for all configurations in Table 1.

\subsection{Further Optimization Using Inline PTX Assembly}
The success of Deepseek-v3 \cite{ref14} has demonstrated that PTX \cite{ref15} can serve as a viable alternative to CUDA for achieving superior performance in GPU computing. Its key advantages include direct hardware control for performance optimization, efficient register allocation for compute-intensive workloads, and the ability to integrate inline assembly for critical kernel optimization. PTX provides developers with fine-grained control over GPU execution while preserving compatibility across NVIDIA's GPU generations. 

To maximize code readability, we selectively converted only the while loop in Algorithm 2 into inline PTX assembly. Our approach involved: (1) generating the complete PTX implementation of the kernel function using the NVCC compiler, and (2) extracting and transforming the while loop section into inline PTX form. However, this initial modification alone did not yield performance improvements, necessitating further optimization of the inline PTX implementation.

PTX code optimization adheres to two key principles: (1) instruction count minimization in one loop and (2) branch operation reduction. Based on these two principles, we optimized three critical aspects of the inline PTX implementation: (1) employing the \textbf{lop3} instruction to compute $\sim$(cur | left |right), reducing the instruction count from three to one; (2) removing two address calculation instructions in the loop, and (3) redesigning the "last-row optimization" logic by using \textbf{selp} instruction to replace two branches. Furthermore, we optimized the computation by substituting the input parameter N with N-1, which removed one redundant subtraction instruction.

\section{Experiment}
The experimental evaluation primarily utilizes three GPU cards: RTX 4090, RTX 5090, and A100. The key hardware specifications \cite{ref11, ref12} of these cards are compared as follows.

\begin{table}[H]
	\centering
	\caption{Hardware Comparison of RTX 4090, RTX 5090, and A100}
\resizebox{\columnwidth}{!}{ 
	\begin{tabular}{|c |>{\centering\arraybackslash}p{1cm}| c|>{\centering\arraybackslash}p{2cm}|>{\centering\arraybackslash}p{2cm}|}
\hline
		\textbf{GPU} & \textbf{CUDA Cores} & \textbf{BandWidth} &   \textbf{Max Shared Memory} &  \textbf{Compute Capability}\\ 
\hline
		RTX 4090 & 16384 & 1008GB/s & 99KB & 8.9 \\ 
\hline
		RTX 5090 & 21760 & 1792GB/s & 99KB & 12.0 \\ 
\hline
		A100 & 6912 & 1935GB/s & 163KB & 8.0 \\
\hline
	\end{tabular}
}
\end{table}

Comparative analysis demonstrates that while the A100 exhibits superior communication bandwidth compared to the RTX 4090, it contains significantly fewer computing cores. In solving the N-Queens problem, we generate numerous mutually independent subproblems through pre-placeing the first few rows. As each subproblem can be processed independently by individual threads - requiring neither inter-thread communication nor CPU interaction - the RTX 4090's greater number of computing cores provides distinct advantages in this computational scenario. Given that the RTX 5090 offers approximately 30\% more computing cores than the RTX 4090, we project a corresponding 30\% improvement in computational performance.

First, we conduct experiments using a single RTX 4090 to validate the effectiveness of our proposed optimization methods. Subsequently, we scale the experiment to 8 GPUs, incorporating both A100 and RTX 5090, to compare the method's performance across different GPU architectures. At last, we will utilize 8 RTX 5090 GPUs to solve the 26-Queens problem.

\subsection{Impact of Pre-placed rows on Speed}

Considering that Configuration 1 offers the largest problem size support capacity (see Section 3.2.1), we selected this configuration to evaluate the performance impact of pre-placing the first R rows on a single RTX 4090. The experimental results are presented in Table 3.

\begin{table}[H]
	\centering
	\caption{Performance Comparison (in milliseconds) of Different N-Queens Problem on a single RTX 4090 with Varying Pre-placed Rows}

\begin{minipage}[c]{1.0\textwidth}
	\begin{tabular}{|c |>{\centering\arraybackslash}p{1.1cm}| >{\centering\arraybackslash}p{1.1cm}|>{\centering\arraybackslash}p{1.1cm}|>{\centering\arraybackslash}p{1.1cm}|>{\centering\arraybackslash}p{1.1cm}|}
\hline
		\textbf{N} & \textbf{R=4} & \textbf{R=5} &   \textbf{R=6} &  \textbf{R=7} &  \textbf{R=8}\\ 
\hline
		17 & 162 & 136 & 142 & 171 & 296 \\ 
\hline
		18 & 343 & 232 & 237 & 314 & 573 \\ 
\hline
		19 & 1564 & 978 & 972 & 1268 & 2524 \\
\hline
		20 & 8963 & 6469 & 6536 & 7495 & 10579 \\
\hline
		21 & 83075 & 52583 & 52501 & 57169 & 66861 \\
\hline
		22 & 630574 & 441022 & 452699 & 486972 & 537077 \\
\hline
	\end{tabular}

\end{minipage}
\end{table}

From Table 3, we can see RTX 4090 demonstrates significant performance sensitivity to the number of pre-placed rows. Optimal result is achieved with 5-row pre-placement, followed by 6-row pre-placement, while other configurations result in notable performance degradation. As problem size increases, computation time exhibits characteristic exponential growth with accelerating rates. This accelerating computational complexity presents significant challenges for larger scale problems.

\subsection{Impact of Block/Stack Size on Speed}
In this experimental section, we fixed the number of pre-placed rows at 6 and utilized a single RTX 4090 GPU to evaluate the performance of the two shared memory optimization approaches (Approach 1 and Approach 2) described in Section 3.2. The experimental results are presented in Table 4.

\begin{table}[H]
	\centering
	\caption{Comparison of N-Queens Solving Times (in milliseconds) Under Different Block and Stack Size Configurations with a Single RTX 4090 GPU}

	\begin{tabular}{|c |>{\centering\arraybackslash}p{0.7cm}| >{\centering\arraybackslash}p{0.7cm}|>{\centering\arraybackslash}p{0.8cm}|>{\centering\arraybackslash}p{0.8cm}|>{\centering\arraybackslash}p{0.8cm}|>{\centering\arraybackslash}p{1cm}|}
\hline
		\textbf{Config} & \textbf{N=17} & \textbf{N=18} &   \textbf{N=19} &  \textbf{N=20} &  \textbf{N=21} & \textbf{N=22}\\ 
\hline
		1 & 142 & 237 & 972 & 6536 & 52501 & 452699 \\ 
\hline
		2 & 137 & 218 & 857 & 5447 & 43355 & 371939 \\ 
\hline
		3 & 133 & 204 & 738 & 4669 & 36990 & 316425 \\
\hline
		6 & 142 & 246 & 988 & 6631 & 53287 & 458433 \\
\hline
	\end{tabular}
\end{table}

The experimental results demonstrate that when employing static shared memory allocation, optimizing stack space utilization to increase block size yields significant performance improvements. In contrast, dynamic shared memory allocation (expanded to 96KB) with block size adjusted to 256 resulted in computational performance     degradation \footnote{Other configurations such as 64KB similarly led to performance deterioration.}. This phenomenon confirms the critical importance of L1 cache - which shares hardware resources with shared memory - for performance optimization. On the other hand, this also indicates that the dynamic allocation mechanism of shared memory can incur performance overhead.

\subsection{Impact of Last-Row Optimization on Speed}
Building upon optimization Approach 3 in Section 3.2, the maximum solvable queens number N can be increased by 1 for each configuration. To evaluate the performance impact of this optimization, we replicated the experimental methodology from Section 4.2. The results are presented in Table 5.

\begin{table}[H]
	\centering
	\caption{Comparison of N-Queens Solving Times (in milliseconds) Under Different Block and Stack Size Configurations with a Single RTX 4090 GPU (With Last-Row Optimization)}

	\begin{tabular}{|c |>{\centering\arraybackslash}p{0.7cm}| >{\centering\arraybackslash}p{0.7cm}|>{\centering\arraybackslash}p{0.8cm}|>{\centering\arraybackslash}p{0.8cm}|>{\centering\arraybackslash}p{0.8cm}|>{\centering\arraybackslash}p{1cm}|}
\hline
		\textbf{Config} & \textbf{N=17} & \textbf{N=18} &   \textbf{N=19} &  \textbf{N=20} &  \textbf{N=21} & \textbf{N=22}\\ 
\hline
		1 & 131 & 219 & 720 & 4606 & 36765 & 316293 \\ 
\hline
		2 & 130 & 211 & 635 & 3951 & 31217 & 267182 \\ 
\hline
		3 & 128 & 196 & 585 & 3472 & 27247 & 233548 \\
\hline
		6 & 131 & 206 & 737 & 4674 & 37121 & 318318 \\
\hline
	\end{tabular}
\end{table}

The data in Table 5 demonstrates that implementing the last-row optimization yields consistent speed improvements compared to the Table 4 baseline. Notably, All Configurations demonstrate a significant speedup of over 25\%. Consequently, this optimization is adopted as the default approach in our study.

Special attention must be paid to implementation efficiency, as inefficient implementation can significantly diminish the potential speed improvements. Algorithm 3 presents two equivalent implementations. The initial implementation realizes the last-row optimization through a separate if-statement, resulting in two consecutive conditional checks that yield negligible speed improvement. By contrast, merging these conditionals into a single if-statement produces significant performance gains.

\begin{table}[H]
	\centering
	\begin{tabular}{l}
		\toprule
		\textbf{Algorithm 3} Two Ways of Last-Row Optimization\\ 
		\midrule
\textbf{Before:} \\
1:\quad    \textbf{if}(valid\_pos == 0) \{  \\
2:\qquad        continue;  \\
3:\quad     \}  \\
4:\quad     \textbf{if}(\_\_popc(cur) == N - 1) \{ \\
5:\qquad        sum++;  \\
6:\qquad        continue;  \\
7:\quad     \}  \\
\textbf{After:} \\
1:\quad     \textbf{if}(valid\_pos == 0 || \_\_popc(cur) == N - 1) \{ \\
2:\qquad        sum += \_\_popc(valid\_pos);  \\
3:\qquad        continue;  \\
4:\quad     \}  \\
		\bottomrule
	\end{tabular}
\end{table}

\subsection{Evaluation of Inline PTX Optimization}
Although only limited optimizations were applied to the inline PTX code, the achieved performance improvements surpassed expectations, as demonstrated in Table 6.
\begin{table}[H]
	\centering
	\caption{Comparison of N-Queens Solving Times (in milliseconds) Using Inline PTX Optimization with a Single RTX 4090 GPU (With Last-Row Optimization)}

	\begin{tabular}{|c |>{\centering\arraybackslash}p{0.7cm}| >{\centering\arraybackslash}p{0.7cm}|>{\centering\arraybackslash}p{0.8cm}|>{\centering\arraybackslash}p{0.8cm}|>{\centering\arraybackslash}p{0.8cm}|>{\centering\arraybackslash}p{1cm}|}
\hline
		\textbf{Config} & \textbf{N=17} & \textbf{N=18} &   \textbf{N=19} &  \textbf{N=20} &  \textbf{N=21} & \textbf{N=22}\\ 
\hline
		1 & 128 & 203 & 658 & 4127 & 32619 & 279885 \\ 
\hline
		2 & 126 & 183 & 609 & 3636 & 28551 & 245090 \\ 
\hline
		3 & 126 & 181 & 566 & 3307 & 25337 & 216740 \\
\hline
	\end{tabular}
\end{table}

Compared with table 5, all configurations achieved a speed\-up exceeding 7\%. These results confirm the viability of assem\-bly level optimizations, heralding PTX optimization as an emerging critical skill in GPU programming.

Although Configuration 3 exhibited the highest computational efficiency, its 6-row pre-placement constraint limited its applicability to N-Queens problems with N$\leq$23. For solving 24-26 Queens problems, we were compelled to adopt Configuration 2, which showed approximately 13\% lower computational efficiency than Configuration 3. When addressing larger problem sizes up to 27 Queens, Configuration 1 became necessary, demonstrating a further 14\% performance reduction compared to Configuration 2.

The three optimization techniques proposed in this work - block/stack size optimization, last-row optimization, and inline PTX optimization - demonstrate consistent performance improvements across various GPU architectures, though the magnitude of gains may differ from those observed on RTX 4090.

\subsection{Multi-GPU Performance Evaluation}
The experimental results above comprehensively validate the effectiveness of our optimization methodology in solving the N-Queens problem. To further enhance computational performance, we expanded our experimental platform from a single GPU card to 8 GPU cards, investigating the efficiency gains through increased parallelism for larger N.

Given the substantial time and resource requirements for multi-GPU experiments with larger N (N>24), optimal computational configurations must be predetermined. We focus on two critical parameters: (1) the impact of pre-placed rows on multi-GPU platform, and (2) the performance differences between uniform and non-uniform task partition strategies.

To validate this hypothesis, we first implemented a uniform partition strategy \footnote{Each GPU was assigned an equal number of subproblems to process.} and adopted configuration 2 to measure per-GPU computation time for 22- and 23-Queens problems with pre-placed rows ranging from 5 to 8 rows under 8 RTX 4090 GPUs. The experimental results (in seconds) are presented in Table 7.

\begin{table}[H]
	\centering
	\caption{Per-GPU Computation Time (ID from 0 to 7) for 22-Queens (Top) and 23-Queens (Bottom) Problems with Uniform Task Partition Under Varying Pre-placed Rows Using 8 RTX 4090 GPUs}

\begin{minipage}[c]{1.0\textwidth}

	\begin{tabular}{|c |>{\centering\arraybackslash}p{0.58cm}| >{\centering\arraybackslash}p{0.58cm}|>{\centering\arraybackslash}p{0.58cm}|>{\centering\arraybackslash}p{0.58cm}|>{\centering\arraybackslash}p{0.58cm}|>{\centering\arraybackslash}p{0.58cm}|>{\centering\arraybackslash}p{0.58cm}|>{\centering\arraybackslash}p{0.5cm}|}
\hline
		\textbf{R} & \textbf{0} & \textbf{1} &   \textbf{2} &  \textbf{3} &  \textbf{4} & \textbf{5} &  \textbf{6} &  \textbf{7} \\ 
\hline
		5 & 28.5 & 32.5 & 33.1 & 41.2 & 41.6 & 41.9 & \textbf{47.8} & 44.4 \\ 
\hline
		6 & 19.1 & 24.0	& 27.2 & 29.1 & 36.1 & 35.6 & \textbf{39.6} & 38.5 \\ 
\hline
		7 & 19.2 &	24.1 & 27.5 & 29.4 & 35.3 & 39.7 & \textbf{42.6} & 42.3 \\
\hline
		8 & 21.5 & 26.4 & 30.3 & 32.5 & 36.2 & 44.5 & 46.4 & \textbf{49.3} \\
\hline
	\end{tabular}

\vspace{1em}

	\begin{tabular}{|c |>{\centering\arraybackslash}p{0.58cm}| >{\centering\arraybackslash}p{0.58cm}|>{\centering\arraybackslash}p{0.58cm}|>{\centering\arraybackslash}p{0.58cm}|>{\centering\arraybackslash}p{0.58cm}|>{\centering\arraybackslash}p{0.58cm}|>{\centering\arraybackslash}p{0.58cm}|>{\centering\arraybackslash}p{0.5cm}|}
\hline
		\textbf{R} & \textbf{0} & \textbf{1} &   \textbf{2} &  \textbf{3} &  \textbf{4} & \textbf{5} &  \textbf{6} &  \textbf{7}\\ 
\hline
		5 & 228 & 272 & 270 & 356 & 332 & 389 & 382 & \textbf{401} \\ 
\hline
		6 & 167 & 216 & 234 & 274 & 313	& 320 & \textbf{353} & 349 \\ 
\hline
		7 & 172 & 218 & 241 & 261 & 338	& 337 & \textbf{389} & 374 \\
\hline
		8 & 184 & 235 & 257 & 269 & 357	& 362 & \textbf{443} & 419 \\
\hline
	\end{tabular}

\end{minipage}
\end{table}

Two principal conclusions emerge from Table 7:
\begin{enumerate}
    \item Optimal performance occurs with 6 row pre-placement, while 5 row and 7 row configurations demonstrate comparable efficiency. Consequently, we fix the pre-placed rows to 6 for solving 20- to 26-Queens problems.
    \item Under uniform task partition, significant variance emer\-ges in per-GPU execution times (exceeding 100\% difference), with systematically longer durations for later-subproblem processing. This observation necessitates adopting a non-uniform partition strategy where higher-numbered GPUs receive progressively smaller task allocations. Further analysis reveals partition dependents on value N. Given the impracticality of customizing allocation schemes for each individual N values, we just implemented a non-uniform partition based on computation time of each GPU in table 7: \textbf{[0.20, 0.15, 0.12, 0.11, 0.11, 0.11, 0.10, 0.10]} for all problems. While potentially suboptimal for specific cases, this scheme delivers substantial performance improvements over uniform partition.
\end{enumerate}

After finalizing the pre-placed rows and task partition ratio, we systematically employed Configuration 2 to benchmark larger N values. Besides RTX 4090, RTX 5090 and A100 are also involved. The execution times for 20- to 25-Queens problems are presented in figure 5. The A100 exhibits substantially inferior computational performance compared to the RTX 4090, showing approximately 45\% performance gap. This conclusively demonstrates that core count is the critical performance determinant for N-Queens problems. The RTX 5090 is approximately 22\% faster than the RTX 4090, falling short of the expected 30\% performance improvement.

\begin{figure}[H]
	\centering
	\includegraphics[width=8cm]{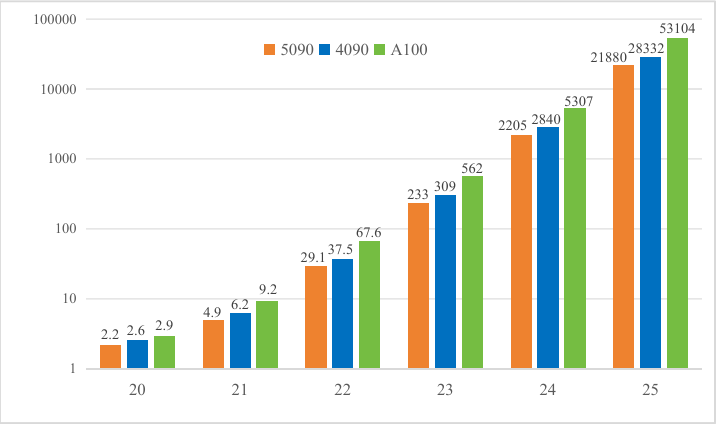}
	\caption{Computation Time (in Seconds) for Solving 20- to 25-Queens Problems on 8 GPU Cards} 
\end{figure}

\begin{table*}[!htbp]
	\centering
	\begin{tabular}{l}
		\toprule
		\textbf{Runtime Log for 27-Queens}\\ 
		\midrule
{[2025-08-18 10:16:52.040] Use 9287.60ms to generate 453688251 subproblems!}\\ 
{[2025-08-18 10:16:53.728] gpu [0] start job, with 90737656(0.20) subproblems.}\\
{[2025-08-18 10:16:53.833] gpu [4] start job, with 49905708(0.11) subproblems.}\\ 
{[2025-08-18 10:16:53.835] gpu [1] start job, with 68053240(0.15) subproblems.}\\ 
{[2025-08-18 10:16:53.838] gpu [7] start job, with 45368815(0.10) subproblems.}\\
{[2025-08-18 10:16:53.842] gpu [5] start job, with 49905708(0.11) subproblems.}\\
{[2025-08-18 10:16:53.847] gpu [6] start job, with 45368828(0.10) subproblems.}\\
{[2025-08-18 10:16:53.853] gpu [2] start job, with 54442588(0.12) subproblems.}\\
{[2025-08-18 10:16:53.860] gpu [3] start job, with 49905708(0.11) subproblems.}\\ 
{[2025-09-10 15:25:18.245] gpu [1] finish job.}\\ 
{[2025-09-10 19:45:56.094] gpu [3] finish job.}\\ 
{[2025-09-10 22:29:02.491] gpu [2] finish job.}\\ 
{[2025-09-12 00:05:49.185] gpu [6] finish job.}\\ 
{[2025-09-12 08:07:13.997] gpu [0] finish job.}\\
{[2025-09-13 05:31:21.437] gpu [4] finish job.}\\ 
{[2025-09-13 08:33:48.516] gpu [7] finish job.}\\
{[2025-09-15 20:15:04.518] gpu [5] finish job.}\\
{[2025-09-15 20:15:07.472] cuda 27 queens result 234907967154122528, calc time: [2455104768.00ms]}\\
		\bottomrule
	\end{tabular}
\end{table*}

Comparing the above figure with Table 6, both cases solve the 22-queens problem using RTX 4090 GPUs, the 8-GPU configuration achieves a speedup of only 6.5× compared to the single-GPU setup (37.5 vs. 245), which falls short of the theoretical expectation. The primary reason is the imperfect load balancing of subproblems, leading to uneven computational workloads across GPUs.

Notably, compared with the state-of-the-art CUDA implementation \cite{ref10}, our solution achieves over 10× performance improvement for the 24-queens problem under same configuration(8 A100 GPUs), and achieves over 26× performance improvement under 8 RTX 5090 GPUs. This breakthrough conclusively validates the effectiveness of our proposed optimization methodology. 

Moreover, we found that for N $\geq$ 23, the computation time for a single problem increases by approximately 10 compared to the N-1 problem, a growth rate far exceeding previous estimates \cite{ref10}. This value is very close to the ratio of the number of solutions for two adjacent N-Queens problems. Therefore, using this ratio to estimate the computation time for larger N-Queens problems is more reasonable.

For the 26-queens problem, we successfully calculated the number of solutions under 8 RTX 5090 GPUs in \textbf{2.5} days. To date, the only known attempt to solve the 26-queens problem was made by Preußer et al., who successfully computed the solution using an FPGA in about 9 months. In contrast, our method achieves the same result over 100 times faster than their approach \footnote{Preußer's result is verified by a supercomputer in about one month, it's too expensive, so just ignore it.}.

For the 27-queens problem, we have two options: Configuration 2 pre-placing queens in first 7 rows, or Configuration 1 pre-placing queens in first 6 rows. We use 24-queens problem to verify which option is faster under 8 RTX 5090 GPUs. Option 1 finishes in 2301 seconds, and option 2 finishes in 2531 seconds. So we  use option 1 to calculate the solution count. The estimated computation time on 8 RTX 5090 GPUs would be approximately 28 days = 2.5 days * 10.52 (27-queens solution count/26-queens solution count) * 1.04 (option 1 is 4\% slower than configuration for solving 24-queens in figure 4). The computation was successfully completed in 28.4 days, a landmark validation that not only conquered the daunting challenge of the 27-queens problem but also demonstrated a remarkable alignment with our initial time projection. The runtime log is shown above. According to the log, there remains a notable disparity in the completion times across GPUs, confirming that the current task partitioning strategy has not yet achieved optimal load balancing for Larger N.

Regarding the unsolved 28-queens problem, the previously leading CUDA implementation was estimated to require over two decades to solve it, while even Preußer's FPGA-based approach would have demanded more than a decade. In contrast, our method can solve 28-queens in approximately 11 months under the 8 RTX 5090 GPU configuration, if scaled to 64 RTX 5090 GPUs, the computation period can reduced to around 50 days. Notably, the 28-queens problem is likely the last instance where the solution count can be stored using the int64 data type. Larger N-queens problems will require support for the int128 data type.

\section{Conclusion}
This paper presents an efficient CUDA-based parallel algorithm for accurately counting solutions to the N-queens problem. By redesigning the implementation of Somers algorithm, we innovatively developed an iterative depth-first search (DFS) version that incorporates extreme stack space compression to fit entirely within GPU shared memory. Notably, we proposed a bank-conflict-free shared memory access pattern that effectively resolves the memory access bottleneck inherent in conventional implementations. Furthermore, we also implemented various optimization techniques to achieve extreme acceleration of the code.

Experimental results demonstrate that our algorithm can solve the 27-queens problem in just 28.4 days using 8 RTX 5090 GPUs, setting a new record for this problem. Compared with lastest CUDA implementation, we achieved over 10x speedup based on same configuration. These provide substantial evidence for the effectiveness of our method. It is worth emphasizing that our implementation not only provides an exemplary case for GPU shared memory optimization, but also exhibits excellent portability and scalability. The concise and efficient code can be easily deployed across various NVIDIA GPU platforms, establishing a reliable performance benchmark for related research.

Although the methodology presented in this paper is applied to the N-Queens problem, the underlying optimization principles are not limited to this specific problem. Any depth-first search problem that satisfies the following conditions can achieve consistent acceleration through our approach:
\begin{enumerate}
	\item It must be capable of generating a sufficiently large number of subproblems;
	\item The recursive algorithm can be transformed into an iterative implementation using a stack structure;
	\item The stack memory required per iteration exceeds one int, as bank conflicts only become possible when the data size exceeds one int.
\end{enumerate}

Examples of applicable problems include sudoku\cite{ref13}, maze searching, combinatorial permutation problems, expression evaluation, and syntactic parsing. Such problems can be effectively accelerated by converting recursion into iteration and leveraging shared memory while avoiding bank conflicts.

However, when addressing larger problem instances (e.g., the 28-queens problem), the current approach still faces challenges with prohibitively long computation times, primarily due to the inherent exponential complexity of the problem. Future research could explore three potential directions for breakthroughs: 1) developing distributed multi-node collaborative computing frameworks; 2) designing robust fault-tolerant recovery mechanisms; 3) investigating more efficient parallel algorithmic paradigms; and 4) use optimization techniques in this paper to accelerate similar problems. We anticipate that future researchers will build upon this work to explore additional efficient solution methods for the N-queens and a series of similar problems, thereby advancing the field's continued progress.

\bibliography{bibfile}


\begin{thebibliography}{15}


\ifx \showCODEN    \undefined \def \showCODEN     #1{\unskip}     \fi
\ifx \showISBNx    \undefined \def \showISBNx     #1{\unskip}     \fi
\ifx \showISBNxiii \undefined \def \showISBNxiii  #1{\unskip}     \fi
\ifx \showISSN     \undefined \def \showISSN      #1{\unskip}     \fi
\ifx \showLCCN     \undefined \def \showLCCN      #1{\unskip}     \fi
\ifx \shownote     \undefined \def \shownote      #1{#1}          \fi
\ifx \showarticletitle \undefined \def \showarticletitle #1{#1}   \fi
\ifx \showURL      \undefined \def \showURL       {\relax}        \fi
\providecommand\bibfield[2]{#2}
\providecommand\bibinfo[2]{#2}
\providecommand\natexlab[1]{#1}
\providecommand\showeprint[2][]{arXiv:#2}

\bibitem[Campbell(1977)]%
        {ref1}
\bibfield{author}{\bibinfo{person}{Paul Campbell}.}
  \bibinfo{year}{1977}\natexlab{}.
\newblock \showarticletitle{Gauss and the eight queens problem: A study in
  miniature of the propagation of historical error}.
\newblock \bibinfo{journal}{\emph{Historia Mathematica}}  \bibinfo{volume}{4}
  (\bibinfo{date}{11} \bibinfo{year}{1977}), \bibinfo{pages}{397--404}.
\newblock
\href{https://doi.org/10.1016/0315-0860(77)90076-3}{doi:\nolinkurl{10.1016/0315-0860(77)90076-3}}


\bibitem[Caromel et~al\mbox{.}(2007)]%
        {ref4}
\bibfield{author}{\bibinfo{person}{Denis Caromel}, \bibinfo{person}{Alexandre
  di Costanzo}, {and} \bibinfo{person}{Clément Mathieu}.}
  \bibinfo{year}{2007}\natexlab{}.
\newblock \showarticletitle{Peer-to-peer for computational grids: mixing
  clusters and desktop machines}.
\newblock \bibinfo{journal}{\emph{Parallel Comput.}} \bibinfo{volume}{33},
  \bibinfo{number}{4} (\bibinfo{year}{2007}), \bibinfo{pages}{275--288}.
\newblock
\showISSN{0167-8191}
\href{https://doi.org/10.1016/j.parco.2007.02.011}{doi:\nolinkurl{10.1016/j.parco.2007.02.011}}


\bibitem[Kise et~al\mbox{.}(2004)]%
        {ref3}
\bibfield{author}{\bibinfo{person}{Kenji Kise}, \bibinfo{person}{Takahiro
  Katagiri}, \bibinfo{person}{Hiroki Honda}, {and} \bibinfo{person}{Toshitsugu
  Yuba}.} \bibinfo{year}{2004}\natexlab{}.
\newblock \showarticletitle{Solving the 24-queens Problem using MPI on a PC
  Cluster}.
\newblock  (\bibinfo{year}{2004}).
\newblock
\urldef\tempurl%
\url{https://api.semanticscholar.org/CorpusID:16371605}
\showURL{%
\tempurl}


\bibitem[Liu et~al\mbox{.}(2024)]%
        {ref14}
\bibfield{author}{\bibinfo{person}{Aixin Liu}, \bibinfo{person}{Bei Feng},
  \bibinfo{person}{Bing Xue}, \bibinfo{person}{Bingxuan Wang},
  \bibinfo{person}{Bochao Wu}, \bibinfo{person}{Chengda Lu},
  \bibinfo{person}{Chenggang Zhao}, \bibinfo{person}{Chengqi Deng},
  \bibinfo{person}{Chenyu Zhang}, \bibinfo{person}{Chong Ruan},
  {et~al\mbox{.}}} \bibinfo{year}{2024}\natexlab{}.
\newblock \showarticletitle{Deepseek-v3 technical report}.
\newblock \bibinfo{journal}{\emph{arXiv preprint arXiv:2412.19437}}
  (\bibinfo{year}{2024}).
\newblock


\bibitem[NVIDIA(2025a)]%
        {ref11}
\bibfield{author}{\bibinfo{person}{NVIDIA}.} \bibinfo{year}{2025}\natexlab{a}.
\newblock \showarticletitle{CUDA C++ Programming Guide}.
\newblock  (\bibinfo{year}{2025}).
\newblock
\urldef\tempurl%
\url{https://docs.nvidia.com/cuda/cuda-c-programming-guide/index.html#features-and-technical-specifications}
\showURL{%
\tempurl}


\bibitem[NVIDIA(2025b)]%
        {ref15}
\bibfield{author}{\bibinfo{person}{NVIDIA}.} \bibinfo{year}{2025}\natexlab{b}.
\newblock \showarticletitle{Parallel Thread Execution ISA Version 9.0}.
\newblock  (\bibinfo{year}{2025}).
\newblock
\urldef\tempurl%
\url{https://docs.nvidia.com/cuda/parallel-thread-execution/index.html}
\showURL{%
\tempurl}


\bibitem[Pantekis et~al\mbox{.}(2024)]%
        {ref10}
\bibfield{author}{\bibinfo{person}{Filippos Pantekis}, \bibinfo{person}{Phillip
  James}, \bibinfo{person}{Oliver Kullmann}, {and} \bibinfo{person}{Liam
  O'Reilly}.} \bibinfo{year}{2024}\natexlab{}.
\newblock \showarticletitle{Optimized massively parallel solving of N‐Queens
  on GPGPUs}.
\newblock \bibinfo{journal}{\emph{Concurrency and Computation: Practice and
  Experience}}  \bibinfo{volume}{36} (\bibinfo{year}{2024}).
\newblock
\urldef\tempurl%
\url{https://api.semanticscholar.org/CorpusID:266836706}
\showURL{%
\tempurl}


\bibitem[Preußer et~al\mbox{.}(2009)]%
        {ref5}
\bibfield{author}{\bibinfo{person}{Thomas Preußer}, \bibinfo{person}{Bernd
  Nägel}, {and} \bibinfo{person}{Rainer Spallek}.}
  \bibinfo{year}{2009}\natexlab{}.
\newblock \showarticletitle{Putting queens in carry chains}.
\newblock \bibinfo{journal}{\emph{Fakultat Informatik, Technische U Niversitat
  Dresden, Tech. Rep}} (\bibinfo{year}{2009}).
\newblock
\showISBNx{TUD-FI09–03}


\bibitem[Preußer(2025)]%
        {ref7}
\bibfield{author}{\bibinfo{person}{Thomas~B Preußer}.}
  \bibinfo{year}{2025}\natexlab{}.
\newblock \showarticletitle{Q27}.
\newblock  (\bibinfo{year}{2025}).
\newblock
\urldef\tempurl%
\url{https://github.com/preusser/q27}
\showURL{%
\tempurl}


\bibitem[Preußer and Engelhardt(2017)]%
        {ref6}
\bibfield{author}{\bibinfo{person}{Thomas~B. Preußer} {and}
  \bibinfo{person}{Matthias~R. Engelhardt}.} \bibinfo{year}{2017}\natexlab{}.
\newblock \showarticletitle{Putting Queens in Carry Chains, No\'{z}27}.
\newblock \bibinfo{journal}{\emph{J. Signal Process. Syst.}}
  \bibinfo{volume}{88}, \bibinfo{number}{2} (\bibinfo{date}{Aug.}
  \bibinfo{year}{2017}), \bibinfo{pages}{185–201}.
\newblock
\showISSN{1939-8018}
\href{https://doi.org/10.1007/s11265-016-1176-8}{doi:\nolinkurl{10.1007/s11265-016-1176-8}}


\bibitem[Sato et~al\mbox{.}(2011)]%
        {ref13}
\bibfield{author}{\bibinfo{person}{Yuji Sato}, \bibinfo{person}{Naohiro
  Hasegawa}, {and} \bibinfo{person}{Mikiko Sato}.}
  \bibinfo{year}{2011}\natexlab{}.
\newblock \showarticletitle{GPU acceleration for Sudoku solution with genetic
  operations}.
\newblock  (\bibinfo{year}{2011}).
\newblock


\bibitem[Somers(2002)]%
        {ref2}
\bibfield{author}{\bibinfo{person}{Jeff Somers}.}
  \bibinfo{year}{2002}\natexlab{}.
\newblock \showarticletitle{The N Queens Problem: a study in optimization}.
\newblock  (\bibinfo{year}{2002}).
\newblock
\urldef\tempurl%
\url{http://users.rcn.com/liusomers/nqueen_demo/nqueens.html}
\showURL{%
\tempurl}


\bibitem[Thouti and Sathe(2012)]%
        {ref9}
\bibfield{author}{\bibinfo{person}{Krishnahari Thouti} {and}
  \bibinfo{person}{S.~R. Sathe}.} \bibinfo{year}{2012}\natexlab{}.
\newblock \showarticletitle{Solving N-Queens problem on GPU architecture using
  OpenCL with special reference to synchronization issues}.
\newblock  (\bibinfo{year}{2012}), \bibinfo{pages}{806--810}.
\newblock
\href{https://doi.org/10.1109/PDGC.2012.6449926}{doi:\nolinkurl{10.1109/PDGC.2012.6449926}}


\bibitem[Wikipedia(2025)]%
        {ref12}
\bibfield{author}{\bibinfo{person}{Wikipedia}.}
  \bibinfo{year}{2025}\natexlab{}.
\newblock \showarticletitle{List of Nvidia graphics processing units}.
\newblock  (\bibinfo{year}{2025}).
\newblock
\urldef\tempurl%
\url{https://en.wikipedia.org/wiki/List_of_Nvidia_graphics_processing_units}
\showURL{%
\tempurl}


\bibitem[Zhang et~al\mbox{.}(2011)]%
        {ref8}
\bibfield{author}{\bibinfo{person}{Tao Zhang}, \bibinfo{person}{Wei Shu}, {and}
  \bibinfo{person}{Min-You Wu}.} \bibinfo{year}{2011}\natexlab{}.
\newblock \showarticletitle{Optimization of N-Queens Solvers on Graphics
  Processors}.
\newblock \bibinfo{journal}{\emph{Advanced Parallel Processing Technologies}}
  (\bibinfo{year}{2011}), \bibinfo{pages}{142--156}.
\newblock
\showISBNx{978-3-642-24151-2}


\end{thebibliography}

\end{document}